\documentclass[aps,prb,twocolumn,showpacs,superscriptaddress]{revtex4}
\usepackage{amsmath}
\usepackage{dcolumn}
\usepackage{epsfig}
\usepackage{graphics}
\usepackage{latexsym}

\def\MGB{MgB$_2$~}

\begin{document}

\hyphenation{Ka-pi-tul-nik}

%\twocolumn[
%\hsize\textwidth\columnwidth\hsize\csname@twocolumnfalse\endcsname

%\draft

\title{Preparation and  properties of amorphous MgB$_2$/MgO superstructures: A new model disordered superconductor}

\author{W. Siemons}
\affiliation{Geballe Laboratory for Advanced Materials, Stanford University, Stanford, California, 94305}
\affiliation{Faculty of Science and Technology and MESA+ Institute for Nanotechnology, University of Twente, P.O. Box 217, 7500 AE, Enschede, The Netherlands}
\author{ M.A. Steiner}
\altaffiliation{Present address: National Renewable Energy Laboratory, Golden, CO 80401}
\affiliation{Geballe Laboratory for Advanced Materials, Stanford University, Stanford, California, 94305}
\affiliation{Department of Applied Physics, Stanford University, Stanford, CA 94305}
\author{G. Koster}
\affiliation{Geballe Laboratory for Advanced Materials, Stanford University, Stanford, California, 94305}
\affiliation{Faculty of Science and Technology and MESA+ Institute for Nanotechnology, University of Twente, P.O. Box 217, 7500 AE, Enschede, The Netherlands}
\author{D.H.A. Blank}
\affiliation{Faculty of Science and Technology and MESA+ Institute for Nanotechnology, University of Twente, P.O. Box 217, 7500 AE, Enschede, The Netherlands}
\author{M.R. Beasley}
\affiliation{Geballe Laboratory for Advanced Materials, Stanford University, Stanford, California, 94305}
\affiliation{Department of Applied Physics, Stanford University, Stanford, CA 94305} 
\author{A. Kapitulnik}
\affiliation{Geballe Laboratory for Advanced Materials, Stanford University, Stanford, California, 94305}
\affiliation{Department of Applied Physics, Stanford University, Stanford, CA 94305} 
\affiliation{Department of Physics, Stanford University, Stanford, CA 94305}

%\email{aharonk@stanford.edu}

\date{\today}

\begin{abstract}
In this paper we introduce a novel method for fabricating MgB$_2$/MgO multilayers and demonstrate the potential for using them as a new model for disordered superconductors. In this approach we control the annealing of the \MGB to yield an interesting new class of disordered (amorphous) superconductors with relatively high transition temperatures. The multilayers appear to exhibit quasi-two-dimensional superconductivity with controlled anisotropy. We discuss the properties of the multilayers as the thickness of the components of the bilayers vary.

\end{abstract}


\maketitle

\section{introduction}

Magnesium diboride (\MGB) has proven to be a difficult material to synthesize in thin film form. \cite{buzea} The primary source of this difficulty lies in the very high vapor pressure of Mg under the conditions of growth. Similar to single crystals, the best method to fabricate high quality thin films of \MGB  has proven to be the Mg diffusion method, which yields results that are almost independent of the substrate.\cite{schlom}  A general approach to the fabrication consists of the deposition of a disordered precursor film, usually with excess Mg, followed by an anneal to form crystalline MgB$_2$. This method will yield high quality films with high transition temperatures. Such films may be important for future device applications such as tunneling devices that work in the range of 25 K, and may have the potential to replace low transition-temperature materials (NbTi, Nb$_3$Sn) in high-field applications. 

Besides its very high transition temperature ($\sim$ 40 K), its low anisotropy ($<$ 6), and very high critical field ($>$ 50 T), \MGB has other properties that make it particularly suitable for studying BCS superconductivity in some very interesting limits. One such property is the two gap feature that is readily observed in high quality crystals. Other examples include the low spin-orbit interaction and this material or the sensitivity to disorder that can be studied over a wider range of disorder due to the initial high transition temperature.  

It is the disorder effect that is the subject of the present paper, where samples have been fabricated in thin film form. These are obtained using a variant of the standard deposition with a post-anneal method in which the annealing is not carried through to completion, thus yielding an interesting new class of disordered (possibly amorphous) superconductors with relatively high transition temperatures. This approach, with a controlled Mg diffusion process, yields a novel method for the fabrication of \MGB/MgO multilayers. The multilayers appear to exhibit quasi-two-dimensional superconductivity and, with further optimization, may provide a new model system for studies of the effects of dimensionality on superconductivity.

\section{samples}

\begin{figure}[h]
\centering
\includegraphics[width=0.8 \columnwidth]{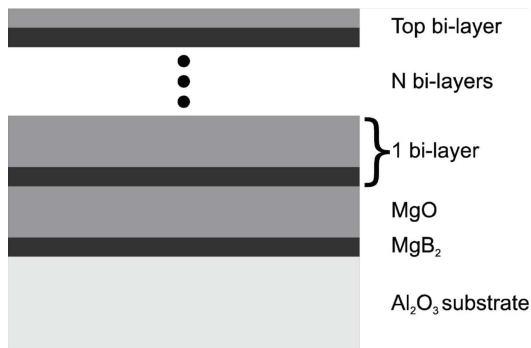}
\vspace{-4mm}
\caption{ A schematic representation of a typical sample. Multilayers were always started with a layer of \MGB and always terminated with a layer of Mg. The magnesium layers were found to be fully oxidized after being exposed to air.} 
\label{layers}
\end{figure}

As it is well known, two  factors complicate the fabrication of \MGB films. These are the high vapor pressure of the magnesium at low temperature and the high sensitivity of magnesium to oxidation, requiring very low oxygen partial pressures in the deposition system. In order to compensate for a possible loss of Mg in the subsequent \emph{in-situ} annealing process, excess Mg was provided during deposition, following the approach of most groups using Pulsed Laser Deposition (PLD).\cite{zeng,blank,christen} This was achieved by successive deposition of bilayers of MgB$_2$/Mg  on a room temperature Al$_2$O$_3$ substrate. The subsequent annealing temperature was kept low in order to obtain disordered films, with lower temperatures (of order $\sim$ 600 $^\circ$C) yielding the most disordered films with a finite superconducting temperature. Subsequent cooldown in the presence of a finite oxygen partial pressure yields a novel multilayer structure which is spontaneously formed with MgB$_2$/MgO being the unit bilayer.  A schematic representation of our samples is shown in Fig.~\ref{layers}. We found that influencing the properties of the superconducting layers in such a layered system was not difficult. 

Films reported in this study were grown by PLD in a stainless steel vacuum chamber with a background pressure of 10$^{-8}$ Torr. A KrF excimer laser produces a 248 nm wavelength beam with typical pulse lengths of 20-30 ns. A rectangular mask shapes the beam, and a variable attenuator permits variation of the pulse energy. The energy density on the targets was kept at approximately 2.3 J/cm$^2$. Before each run the targets were pre-ablated for one minute at 20 Hertz each.

The multilayers were grown on Al$_2$O$_3$ R-plane substrates, which were annealed in an oxygen atmosphere at 1100 $^\circ$C for one hour to create a high quality surface. Substrates were glued onto the stainless steel sample holder with silver paste and cured for ten minutes at 150 $^\circ$C before the chamber was evacuated. The samples were grown by alternating between targets of sintered \MGB and metallic Mg; each bilayer was deposited by ablating first \MGB~and then Mg. The last layer in the structure was made thinner than the others to enable good electrical connection to the conducting layer. The properties of the samples, for example $T_c$, and the coupling between superconducting layers, could be influenced by varying the thicknesses of the \MGB and MgO layers. A shadow mask was used to create a standard transport-Hall bridge pattern in order to facilitate electrical transport measurements.

The MgB$_2$/Mg layers were deposited at room temperature and subsequently annealed \emph{in-situ} in 100 mTorr flowing argon at 600 $^\circ$C for one hour. The samples were then cooled down and removed from the chamber, and the excess Mg oxidized. Care was taken to minimize the time between fabrication and measurement, to avoid degradation of the properties because of the exposure to air.

\begin{figure}[h]
\centering
\includegraphics[width=0.9 \columnwidth]{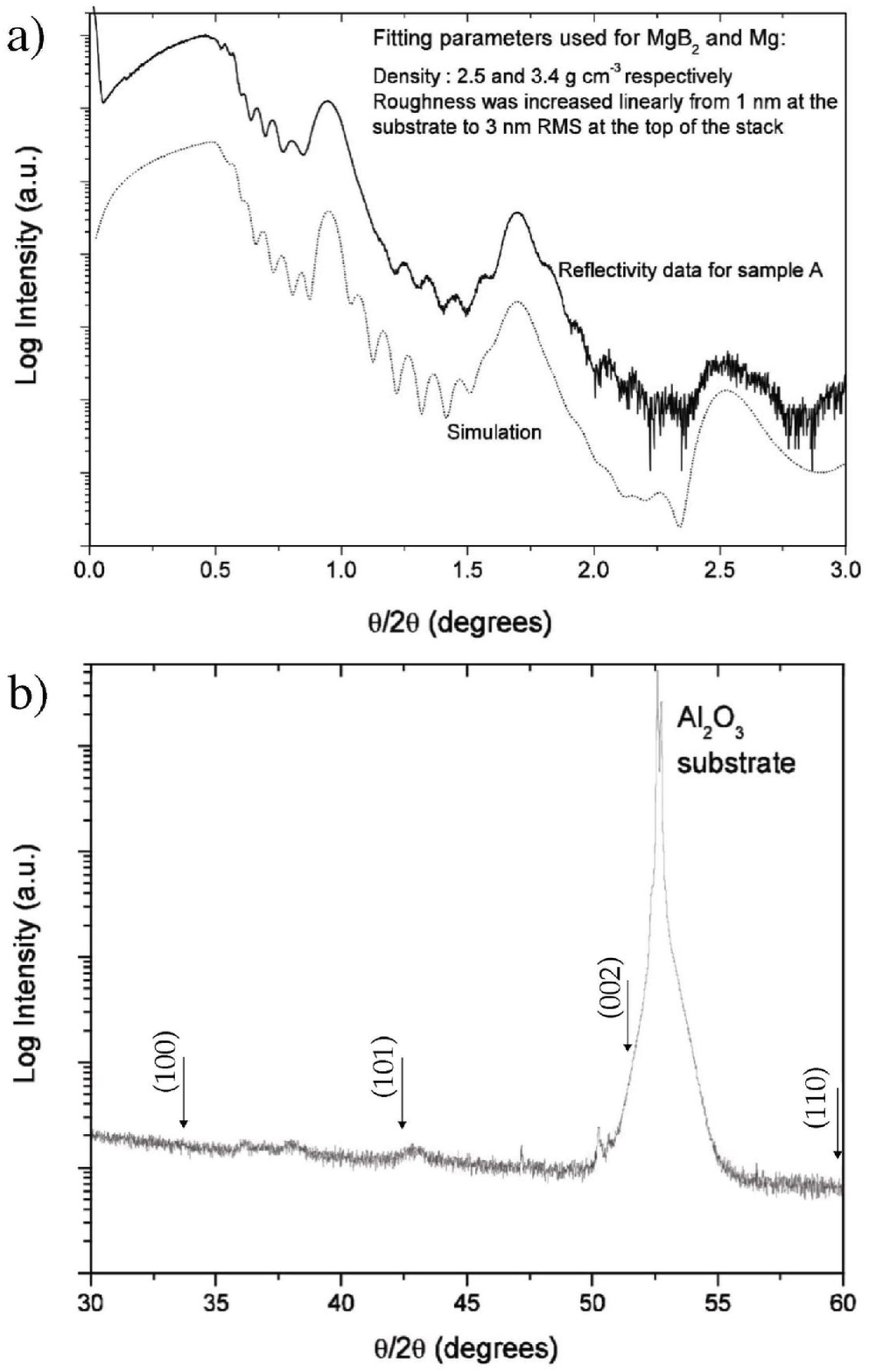}
\vspace{-4mm}
\caption{ a) X-ray reflectivity measurement of sample A. Two periods are visible, indicative of a sample with two different types of interfaces. The best simulation (dotted line) was obtained when only 7 bilayers were included in the model. The densities of the layers were close to the theoretical values of \MGB and MgO and the roughness was increased linearly from the substrate to the top surface. The simulation is offset for clarity. b) High angle XRD measurement of sample A. The main peak is due to the  Al$_2$O$_3$ substrate. Arrows point to where \MGB lines are expected.} 
\label{xray}
\end{figure}

Both low-angle (reflectivity) and high-angle X-Ray diffraction (XRD) experiments were performed on a Philips X'pert thin film diffractometer. For the reflectivity measurements unpatterned samples were used to maximize the signal. Electrical transport properties were measured with a Quantum Design PPMS system, in a magnetic field up to nine Tesla. Transport measurements at temperatures below 1 K were carried out in an Oxford 400 dilution refrigerator. The linear sheet resistance was measured using standard four-point lockin techniques at frequencies from 3 Hz to 17 Hz, with excitation currents kept below 1 nA. Transport measurements results will be discussed in the next section. The samples reported here are listed in Table~\ref{table} along with some of their properties. 

\begin{table}[h]
\caption{Characteristics of samples reported in this paper. Multilayers consist of 8 bilayers of \MGB/MgO.}
\label{table}
\begin{center}
\begin{tabular}{|c|l|l|l|l|l|l|l|}
\hline
Sample & MgB$_2$ & Mg  & d-MgB$_2$  & d-Mg  & d-film &  R$_\Box$(40K) & T$_{c}$ \\
 & pulses & pulses &  [nm] & [nm] & [nm] & [$\Omega$]  &  [K]\\ \hline \hline
A & 360 & 120 & 8.0 & 3.0 & 77 & 91.3 & 24\\ \hline
B & 360 & 600 & 8.0 & 15 & 185 & 46 & 17\\ \hline
C & 240 & 600 & 5.3 & 15 & 163 & 786 & 9\\ \hline
D & 220 & 600 & 4.9 & 15 & 160 & 1363 & 7\\ \hline
E & 180 & 600 & 4.0 & 15 & 153 & 420000 & -\\ \hline
%F & 360 & 120 & 8.0 & 3 & 77 & 91.3 & 24\\ \hline
\end{tabular}
\end{center}
\end{table}

Figure~\ref{xray} shows a typical low-angle x-ray diffraction measurement on one of the samples. This particular sample was deposited using 360 pulses of \MGB and 120 pulses of Mg for every bilayer (sample A in Table~\ref{table}), repeating this sequence eight times. Apparent in Fig.~\ref{xray} are oscillations of two distinct periods. We interpret these oscillations in terms of a superstructure consisting of repeating bilayers. The oscillations with the large period come from the layered pairs, and those with the short period come from the entire film. 

From the period of the large peaks, we estimate the thickness of a bilayer to be approximately 11 nm. Since the film was created by alternately pulsing between \MGB and Mg eight times, one would expect that there are eight repeats of the bilayers. However, the best simulation (the dotted line in Fig.~\ref{xray}a) was obtained using only seven such layers, resulting in a total film thickness of 77 nm. We believe this discrepancy in the total number of layers is due to a roughening of the superstructure that occurs with increasing thickness, causing the top layer to be destroyed and not show up in reflectivity measurements. SEM measurements support this roughening conjecture and show a relatively flat template on which large (presumably boron\cite{li}) particles are present. The simulation of the reflectivity measurements also tells us that the densities of the materials are very close to the theoretical values expected for \MGB and MgO, indicating that the magnesium in the MgO layers has fully oxidized. To get a good simulation for the higher angles ($> 1^\circ$) the RMS roughness was increased linearly from the substrate to the top surface.

High angle XRD measurements show the substrate peak and some very broad low intensity peaks that can be attributed to MgO, as is shown in Fig.~\ref{xray}b. The lack of strong \MGB peaks confirms that the \MGB is highly disordered and possibly even amorphous.

\section{normal state transport and the superconducting transition}

The normal state resistance  is a first good indicator of the disorder in our samples. For high quality single crystals the resistivity has been found to drop with temperature, exhibiting a temperature dependence $\Delta \rho \propto T^\alpha$ with $\alpha > 2.5$ and resistivity ratio exceeding  25.\cite{finnemore,canfield,xu,kazakov} Similar behavior was also found for  high quality, dense wires of MgB$_2$.\cite{canfield} Quality thin films, usually grown epitaxially, exhibit similar normal state properties.\cite{zeng} However, upon introduction of disorder, single crystals show a diminished resistivity ratio. For carbon-doped crystals this ratio was reduced to order unity, while the superconducting transition temperature decreased monotonically with increasing carbon content.\cite{kazakov}

\begin{figure}[h]
\begin{center}
\includegraphics[width=0.9 \columnwidth]{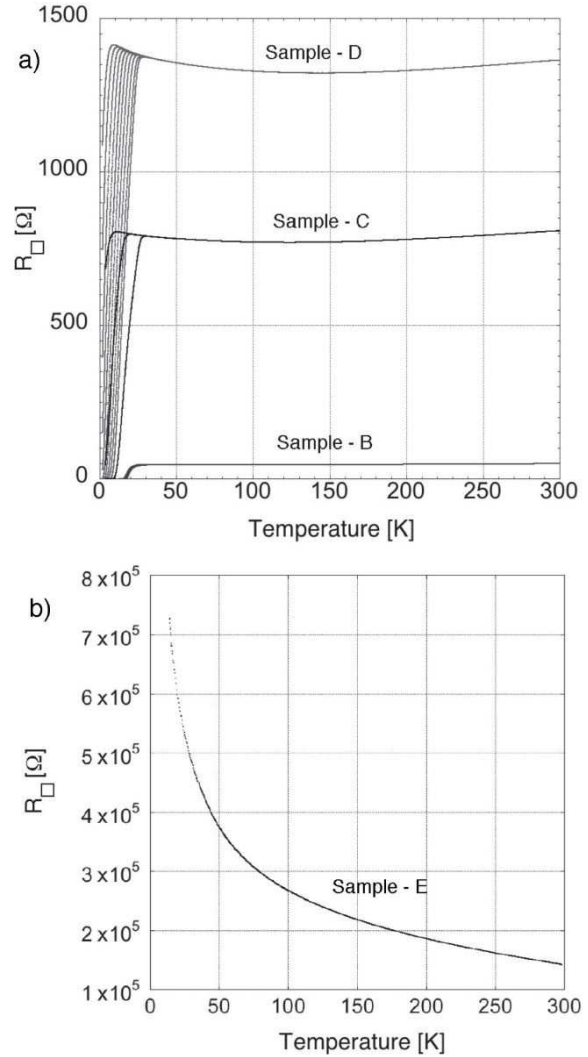}
\end{center}
\vspace{-4mm}
\caption{ Normal state resistivity of samples B, C, D (top) and E (bottom). Note the different scale for the two panels. For the top panel we also show the transitions in a magnetic field. These are discussed in the text.} 
\label{normal}
\end{figure}

Inspecting the normal state resistance of our films in Fig.~\ref{normal}, we observe that the samples all show very low resistivity ratios, similar to disordered bulk materials. Note that we have calculated the two-dimensional resistivity, $R_\Box$, for the full thickness of the sample and not per layer. The resistance per square for films A (not shown) and B is low even though the resistivity is of order 10$^{-3}$ $\Omega$-cm, which no doubt explains the almost flat temperature dependence of the resistance. The transition temperature $T_c$ of these two samples is reduced by a factor of 2-3 from that of bulk, high-quality crystals $T_c$. Samples C and D are more anisotropic and consequently show larger resistance. These two samples also show a resistance minimum at $\sim$150 K, signaling the onset of localization processes. Indeed, a slight change in the thickness of the \MGB superconducting layers, from 4.9 to 4.0 nm, drives the sample from being metallic and then superconducting (sample D) to being insulating (sample E), despite the same MgO spacer layers and same total number of layers. 
 
In Fig.~\ref{normal}a we plotted the normal state resistance in magnetic fields between zero and 9 T, visible on the left side of the graph.  While we examine the superconducting transitions in a magnetic field in greater detail in the next section, we note here the rather simple behavior of the transitions in low fields as being simply superconducting pair-amplitude controlled. This is clearly seen by the fact that the normal state resistance seems to continue smoothly to lower temperatures with increasing field.

\begin{figure}[h]
\begin{center}
\includegraphics[width=0.9 \columnwidth]{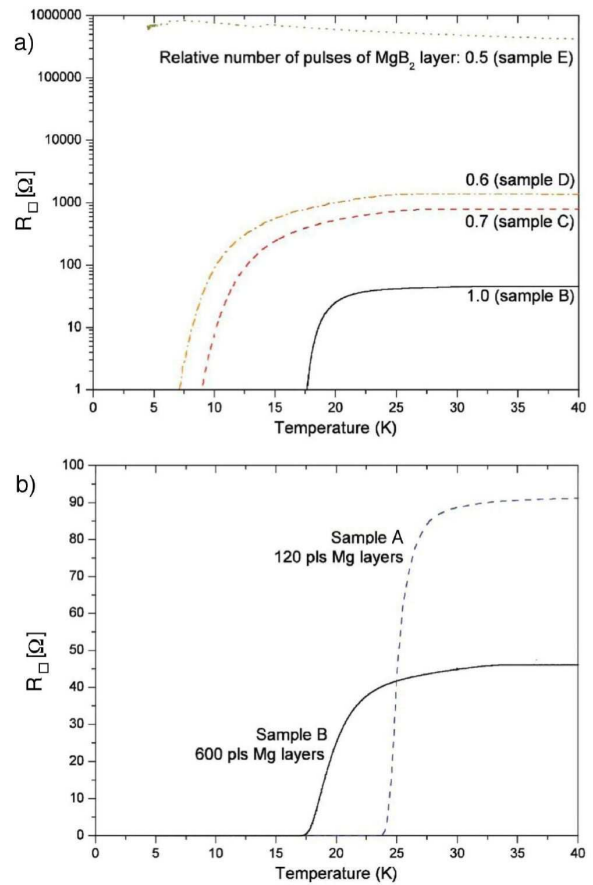}
\end{center}
\vspace{-4mm}
\caption{ a) The zero-field resistance for different \MGB superconducting layer thicknesses, keeping the MgO separating layer constant at 600 pulses. Note the logarithmic scale. From bottom to top the curves of samples B,C,D and E are plotted. The number of pulses for the \MGB layer is given for each curve relative to the 360 used for sample B.  b) Comparison of samples A and B with 360 pulses of \MGB  but different number of Mg pulses, resulting in a different spacing between the superconducting layers. The more decoupled sample (A) shows the greater reduction in $T_c$, down to 17 K, compared to 24 K for the other sample. } 
\label{samples}
\end{figure}

Figure~\ref{samples} concentrates on the zero-field resistive transitions of the representative samples listed in Table~\ref{table}. Fig.~\ref{samples}a shows the transitions for various thicknesses of the \MGB layers, with a fixed MgO thickness, while Fig.~\ref{samples}b shows the transitions at various MgO thicknesses, with a fixed \MGB thickness. Two prominent features of these data are that the normal state sheet resistance is roughly independent of temperature and that $T_c$ is reduced from the bulk value of 39 K, both as expected for disordered superconductors. Nevertheless, in absolute terms the $T_c$'s are distinctly high for a disordered conventional superconductor.  Also evident from Fig.~\ref{samples}a is that the normal state sheet resistance increases and $T_c$ decreases as the \MGB layer is made thinner. This may be attributed to the stronger two-dimensionality and thus weaker superconductivity of the \MGB layers, unable to strengthen due to the thick, intervening MgO layers. This interpretation is reinforced by Fig.~\ref{samples}b, showing that $T_c$ decreases as the thickness of the MgO layers increases, further decoupling the \MGB layers.  

\section{Upper critical field}

The best way to study the coupling between layers is by using a magnetic field.  Indeed, magnetic properties of high-temperature superconductors, especially in the vortex state, revealed much of their anisotropic character.\cite{ando,bible}   This reality led White {\it et al.} to propose that artificial multilayers of MoGe/Ge could be used as a tunable model system for the study of the magnetic properties of layered superconductors.\cite{white}

\begin{figure}[h]
\begin{center}
\includegraphics[width=0.9 \columnwidth]{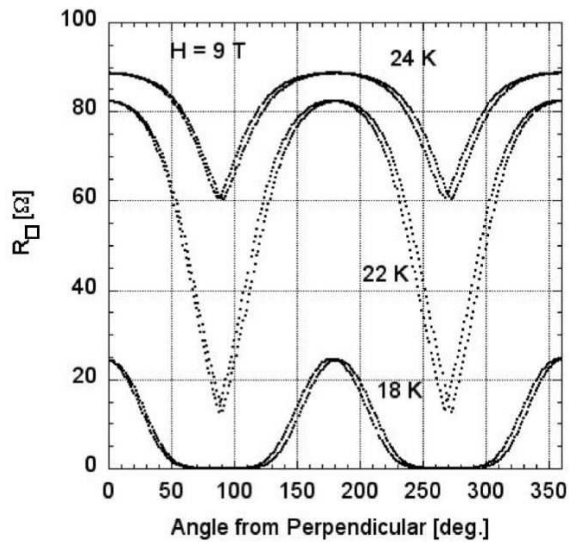}
\end{center}
\vspace{-4mm}
\caption{ Angular dependence of the resistance at $H = $9 T for sample A. The strong anisotropy is evident in the 18 K data.} 
\label{angle}
\end{figure}

The anisotropy of the multilayers is demonstrated through the angular dependence of the resistance near and below $T_c$. Figure~\ref{angle} shows the strong anisotropy of sample A, as the temperature is lowered below the zero-field $T_c$ at an applied field of 9 T. This sample, with the highest $T_c$ in Table~\ref{table}, has relatively strong anisotropy demonstrated by the fact that a few degrees below $T_c$ the parallel field resistance is zero while the perpendicular one is a significant fraction of the normal state resistance.

\begin{figure}[h]
\begin{center}
\includegraphics[width=0.9 \columnwidth]{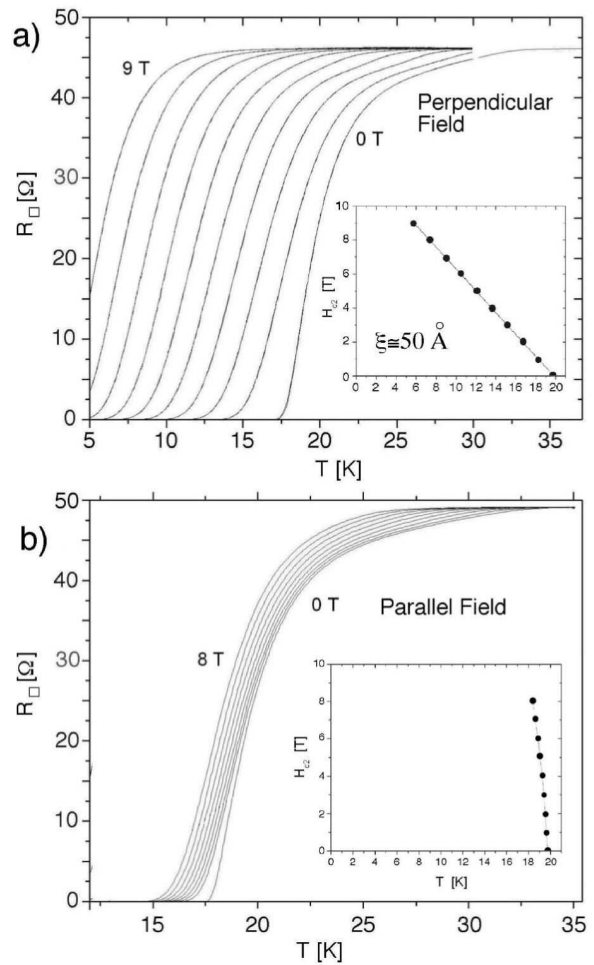}
\end{center}
\vspace{-4mm}
\caption{  a) Resistive transitions of sample B in a perpendicular magnetic field. The fields range from 0 to 9 T, in increments of 1 T. The inset shows the dependence on transition temperature of the in-plane critical magnetic field, $H_{c2}$ (taken at 50$\%$ of the transition). The coherence length was calculated to be 51 $\AA$. b) Resistive transitions in a parallel magnetic field. The inset shows the critical magnetic field plotted against the transition temperature at 50$\%$ of the transition. } 
\label{field1}
\end{figure}

Typical results for the magnetic field dependence of the resistive transition and the associated upper critical field curves $H_{c2}(T)$ (perpendicular and parallel to the layers) are shown in Fig.~\ref{field1} for sample B. A set of resistive transitions in perpendicular fields up to 9 T shows broad transitions that are roughly parallel with increasing field. From this data we can determine the in-plane coherence length. Choosing a mid-point of the transition, the perpendicular critical field (see inset in Fig.~\ref{field1}a) shows a linear behavior with temperature, near $T_c$. From the slope of the critical field near $T_c$ we calculate the zero-temperature coherence length using: $-T_c(dH_{c2\perp}/dT)=\Phi_0/(2\pi \xi^2)$. We find that the in-plane zero-temperature Ginzburg-Landau (GL) coherence length $\xi_{in} \approx 50 \AA$, about 50$\%$ lower than the in-plane coherence length found for high quality bulk samples of MgB$_2$.\cite{finnemore}  Resistive transitions in a parallel field also shows parallel curves, with $T_c$ decreasing as the field increases. Again we choose a mid-point of the transition and plot the parallel critical field in the inset of Fig.~\ref{field1}b.  

For this parallel-field case we note that two anisotropy effects need to be considered. The first is the inherent anisotropy of the \MGB material which is of order 6 for high quality crystals but was found to be substantially reduced when disorder is introduced.\cite{eisterer} Indeed for disordered crystals\cite{eisterer} or films,\cite{gandikota} the inherent anisotropy between the $ab$ and $c$ directions of the \MGB can be reduced to unity. The second effect is the anisotropy due to the layered nature of our sample.\cite{ruggiero} The very steep increase of the parallel field near $T_c$ prevents us from determining the coherence length in that direction. We conclude that this coherence length has to be much smaller than the single \MGB layer thickness, and the mass-anisotropy ratio $\propto (\xi_{in}/\xi_{out})^2 >> 1$. The large anisotropy observed is therefore a direct result of the layered nature of our sample and not the inherent anisotropy of the \MGB layers.

\begin{figure}[h]
\begin{center}
\includegraphics[width=0.9 \columnwidth]{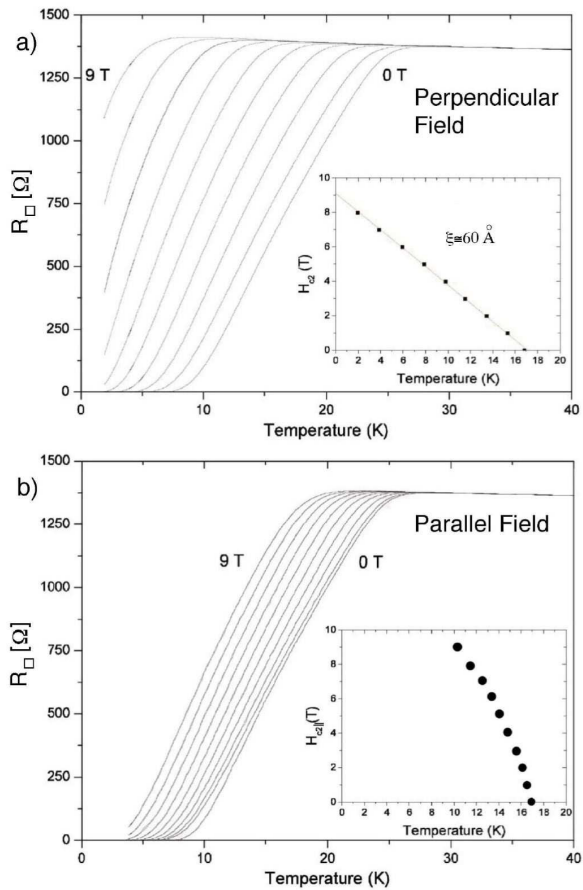}
\end{center}
\vspace{-4mm}
\caption{  a) Resistive transitions of sample D in a perpendicular magnetic field. As in Figure 6, the fields range from 0 to 9 T, in increments of 1 T. The inset shows the dependence on transition temperature of the in-plane critical magnetic field, $H_{c2}$ (taken at 50$\%$ of the transition). The coherence length was calculated to be 58 $\AA$. b)Resistive transitions in a parallel magnetic field. The inset shows the critical magnetic field plotted against the transition temperature at 50$\%$ of the transition. } 
\label{field}
\end{figure}

We next turn to sample D which is more disordered than sample B based on its normal state resistivity.  Resistive transitions in a field for sample D are shown in Figure~\ref{field}.  Compared with the transitions for sample B, those of sample D are significantly broader. However, since the transitions are still roughly parallel with increasing field, we can again determine in-plane coherence length from the slope of $H_{c2}(T)$ versus $T_{c}$. Choosing again a mid-point of the transition, the perpendicular critical field shows a linear behavior near $T_c$, and from the slope near $T_c$ we calculate  $\xi_{in} \approx 60 \AA$.

For the parallel field transitions, we also choose a mid-point of the transition for the calculation of coherence lengths.  The inset in Fig.~\ref{field}b  shows a strong curvature behavior below $T_c$ characteristic of  two-dimensional systems. The behavior appears less steep than for sample B and thus we attempt to analyze the parallel critical field behavior as a crossover from three to two dimensions, for which the critical field near $T_c$ has the temperature dependence $H_{c2||}\propto (T_c - T)^{1/2}$. In the limit as $T \rightarrow T_c$, one expects the three-dimensional result $H_{c2}\propto (T_c - T)$. For a multilayered system such behavior is a signature of decoupled layers.\cite{ruggiero} Indeed, close inspection of the behavior near $T_c$ suggests that below $\sim 2$ T a straight line can be fit to the data with a slope of $\sim 2.4$ T/K  yielding a coherence length perpendicular to the planes of $\xi_{out} \approx 13 \AA$, smaller than the superconducting layer thickness. The ratio of the two coherence lengths is $\sim$ 4.6 (implying a mass-anisotropy value of $\sim$ 20) which, as explained above, should be viewed as the anisotropy of the multilayered system rather than the intrinsic anisotropy of MgB$_2$.

The pure two-dimensional regime of the parallel critical field can be obtained by plotting the square of the parallel critical field versus temperature. For the data in Fig.~\ref{field}b, a linear regime is obtained below $\sim$ 14 K. In this regime we can use the expression\cite{tinkham} $H_{c2 ||}(0)=\Phi_0/(2\pi  \xi_{ab}\tilde{d})$, where $\tilde{d}$ is some effective thickness of order the superconducting layer thickness. Using the data, we obtain $\tilde{d} \approx 33 \AA$, a value that is  close to the thickness of the superconducting layer ($49 \AA$, see table~\ref{table}). We note however that  the data do not extrapolate to the measured $T_c$. One possible reason for the discrepancy is that since the two-dimensional regime occurs when the interlayer coherence length decreases below the bilayer thickness the lower $T_c$ may point to the interaction-reduced $T_c$ typical for thin superconducting films.\cite{mef} 
           
\section{Low temperatures and high fields}

\begin{figure}[h]
\begin{center}
\includegraphics[width=0.9 \columnwidth]{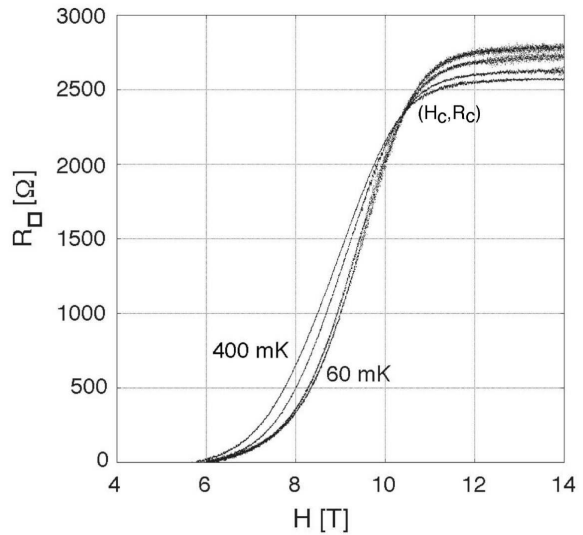}
\end{center}
\vspace{-4mm}
\caption{Resistance of sample D as a function of perpendicular magnetic field for 400 mK, 200 mK, 120 mK and 60 mK. The curves show a crossing point at $H_c =$ 10.42 T and $R_c =2350 ~\Omega/\Box$.} 
\label{scaling}
\end{figure}

We turn now to the analysis of behavior of the multilayers at very low temperatures and high magnetic fields. Figure~\ref{scaling} shows a series of isotherms of the resistance as a function of magnetic field, for sample D. The isotherms each increase in resistance above some characteristic field, and appear to saturate at high field. The value of this saturation resistance increases with decreasing temperature, which is an indication of increased localization effects with lower temperature. Another interesting feature of the data is that the isotherms all cross at a point ($H_c, R_c$). In general, such a crossing is a signature of a zero-temperature quantum phase transition for which the magnetic field is a tuning parameter.

Indeed, disordered superconducting films are expected to undergo a phase transition to an insulating state in the limit of zero temperature, tuned by either a magnetic field or by strong disorder.\cite{goldman} In the case of the field-tuned transition, one finds that the linear resistance decreases with decreasing temperature when the field is below a critical field $H_c$, and increases with decreasing temperature when the field is above the critical field. However, the insulating behavior observed in our films above the crossing point is very weak, and in fact similar to the behavior observed in high temperature superconductors as was previously described by Steiner {\it et al.}\cite{steiner}

The crossing point observed in Fig.~\ref{scaling} also relates to the determination of the upper critical field for this sample. Note that in the previous section, using the mid-point of the transition we found an almost linear dependence of $H_{c2}(T)$ all the way to zero temperature, extrapolating to $H_{c2}(0) \approx 9.2 $T. This value is significantly less than the crossing point field of $H_c$ = 10.42 T. If, instead of using the mid-point we use the actual zero resistance point of the sample to determine $H_{c2}$, we get an initial slope of 0.68 T/K (implying an in-plane coherence length of 87 $\AA$), which appears to become very steep below $\sim$ 1.5 K. Similar behavior has been observed previously for the resistance of high-$T_c$ superconductors with large anisotropy.\cite{mackenzie} This behavior is consistent with the high field crossing point and, since \MGB is believed to be a conventional BCS superconductor, this suggests that the anomalous behavior of $H_{c2}$ is a property of the quasi-two-dimensional layered nature of the materials rather than the mechanism for high-$T_c$. Indeed Kotliar and Varma \cite{kotliar} have suggested that the anomalous behavior of $H_{c2}$ near zero temperature observed in some high-temperature superconductors is a consequence of the proximity to a zero-temperature quantum critical point, which is the zero-temperature critical point at the second-order end point of the first-order melting line of the vortex lattice. Shimshoni {\it et al.}\cite{shimshoni} made a similar connection to the superconductor-insulator transition in thin disordered films of MoGe.\cite{ephron,mk1}

\section{conclusions}

In this paper we demonstrated a new method for producing multilayers of \MGB with MgO as the insulating spacing layer. We have shown that the anisotropy of the multilayers can be controlled with the spacing of the insulating layer, while $T_c$ is a function of the disorder in the individual layers and the coupling between layers. We have also shown that these materials exhibit all of the characteristic features of disordered superconductors, including a signature of a superconductor-insulator transition. We hope that these multilayer structures could be used as a model system for a future study of disordered quasi-two-dimensional superconductors.

\acknowledgments
Author G.K. thanks the Netherlands Organization for Scientific Research (NWO, VENI). W.S. thanks the Nanotechnology network in the Netherlands, NanoNed. We would like to thank Prof. Sang-Im Yoo  for providing us with the \MGB~target used in the PLD. Work supported by NSF-DMR-04 06339 and by DoE contract DE-AC02-76SF00515.

\end{document}